\documentclass[twocolumn,showpacs,preprintnumbers,amsmath,amssymb]{revtex4}

\usepackage{graphicx}
\usepackage{dcolumn}
\usepackage{bm}

\begin{document} 


\title{The one-way CNOT simulation}

\author{Paola Campanale\dag, Domenico Picca\dag * \\\dag Physics Department, University of Study of Bari, Italy\\ *I.N.F.N section of Bari, Italy \\paolacampanale@yahoo.it, picca@ba.infn.it}

\date{\today}

\begin{abstract}
In this paper we present the complete simulation of the quantum
logic CNOT gate in the one-way model, that consists entirely of
one-qubit measurements on a particular class of entangled states.
\end{abstract}

\pacs{03.67.Lx, 03.65.Ud}
\maketitle

\section{Introduction}
The one-way quantum computation model provides a new and
alternative way to build quantum computer, more efficiently than
the logic network model \cite{1},\cite{2}. The quantum gates
simulation is operated by highly entangled cluster states of a
large number of qubits. The entangled state of the cluster serves
as a universal ``substratum'' for any quantum gate. To implement
unitary transformation it suffices to make only one-qubit
measurements. The measurement type and the order in which the
measurements are performed, determine the implemented quantum
algorithm. The calculations results are written onto the
``output'' qubits of the cluster, which are thereby the quantum
register. The output qubits are read with one-qubit measurements
too, by which the classical readout is obtained. The result of
previous measurements determine in which basis the output qubits
need to be measured for the final readout, or, if the readout
measurements are in the $\sigma_{x},\sigma_{y},\sigma_{z}$
eigenbasis, how the readout measurements have to be interpreted.
The individual measurement results are random but correlated.
These correlations enable the quantum computation.

The entangled cluster states can be created efficiently in any
quantum system with a quantum Ising-type interaction between
two-state particles in a lattice configuration at very low
temperatures. Experimentally, the cluster states can be created as
follows. First, a product state:

\begin{eqnarray}
\mid+>_{C}=\bigotimes_{a\in C}\mid+>_{a}
\end{eqnarray}
where ``a'' is the qubit number of the cluster, is prepared
putting all qubits in the $\mid +>$ state.

Second, the unitary transformation $S^{(C)}$:

\begin{eqnarray}
S^{(C)}=\prod_{a,b\in C\mid b-a \in \gamma_{f}}S^{ab}
\end{eqnarray}
is applied to the state $\mid+>_{C}$ . The letter ``f'' shows the
cluster dimension. For the cases of dimension $f=1, 2, 3$ we have:

\begin{eqnarray}
\gamma_{1} & = & \{1\} \nonumber \\
\gamma_{2} & = & \{(1,0)^{T}, (0,1)^{T}\} \nonumber \\
\gamma_{3} & = & \{(1,0,0)^{T}, (0,1,0)^{T}, (0,0,1)^{T}\}.
\end{eqnarray}

The two-qubit transformation $S^{(ab)}$ is such that the state
$\mid1>_{a}\otimes \mid1>_{b}$ acquires a phase of $\pi$ under its
action whereas the remaining states

 $$\mid0>_{a}\otimes
 \mid0>_{b},\hspace{3mm} \mid0>_{a}\otimes \mid1>_{b},\hspace{3mm} \mid1>_{a}\otimes
 \mid0>_{b}$$

acquire no phase. Thus, $S^{(ab)}$ has the form:

\begin{eqnarray}
S^{(ab)} & =& \frac{1}{2}(\textbf{1}^{(a)}\otimes\textbf{1}^{(b)}+
{\sigma^{(a)}}_{z}\otimes\textbf{1}^{(b)} \nonumber \\&+&
\textbf{1}^{(a)}\otimes{\sigma^{(b)}}_{z} -
{\sigma^{(a)}}_{z}\otimes {\sigma^{(b)}}_{z}).
\end{eqnarray}

The unitary transformation $S^{(C)}$ thereby acts only onto the
near neighboring qubits .

The cluster state $\mid+>_{C}$ obeys the eigenvalue equations:

\begin{eqnarray}
\mid\phi>_{C}=
S{\sigma^{(a)}}_{x}S^{\dag}\mid\phi>_{C},\hspace{2mm} \forall a
\in C.
\end{eqnarray}
where, for brevity, $S=S^{C}$.

To obtain $S{\sigma^{(a)}}_{x}S^{\dag}$, one observes that:

\begin{eqnarray}
S^{(ab)}{\sigma^{(a)}}_{x}{S^{(ab)}}^{\dag}\mid\phi>_{C} & = &
{\sigma^{(a)}}_{x}\otimes
{\sigma^{(b)}}_{z}\mid\phi>_{C} \nonumber \\
S^{(ab)}{\sigma^{(a)}}_{x}{S^{(ab)}}^{\dag}\mid\phi>_{C} & = &
{\sigma^{(a)}}_{z}\otimes {\sigma^{(b)}}_{x}\mid\phi>_{C}
\end{eqnarray}

and

\begin{eqnarray}
S^{(ab)}{\sigma^{(c)}}_{x}{S^{ab}}^{\dag}\mid\phi>_{C}={\sigma^{(c)}}_{x}\mid\phi>_{C},\hspace{2mm}
\forall c\in C\backslash\{a,b\}.
\end{eqnarray}

Furthermore for the Pauli phase flip operators $\sigma_{z}$:

\begin{eqnarray}
S^{(ab)}{\sigma^{(d)}}_{z}{S^{(ab)}}^{\dag}={\sigma^{(d)}}_{z},\hspace{2mm}
\forall d\in C\backslash\{a,b\}.
\end{eqnarray}

From the preceding equations, one gets:

\begin{eqnarray}
S{\sigma^{(a)}}_{x}S^{\dag} = {\sigma^{(a)}}_{x}\bigotimes_{b\in
nbgh(a)}{\sigma^{(b)}}_{z}.
\end{eqnarray}

Thus the cluster states $\mid\phi>_{C}$ generated from
$\mid+>_{C}$ via the transformation $S^{(C)}$ obeys the eigenvalue
equations:

\begin{eqnarray}
{K^{a}\mid\phi_{(\kappa)}>_{C}= \mid\phi_{(\kappa)}>_{C}}
\end{eqnarray}

with

\begin{eqnarray}
{K^{a} = {\sigma_{x}}^{a}\bigotimes_{b\in nbgh(a)}
{\sigma_{z}}^{b}}.
\end{eqnarray}

The CNOT gate is a controlled gate with two input named control
and target respectively. The cluster structure which is needed for
the CNOT gate is showed in fig.1. The cluster is composed of 15
qubits. The qubits 1 e 9 are the input qubits of the gate, the
qubits 7 e 15 are the output qubits. The gate simulation is
obtained in two step:

\begin{enumerate}
\item entangled state formation (section III); \item measurements
onto the cluster (section IV). \end{enumerate}

\begin{figure}[htbp]
    \begin{center}
    \includegraphics[width=6.5cm]{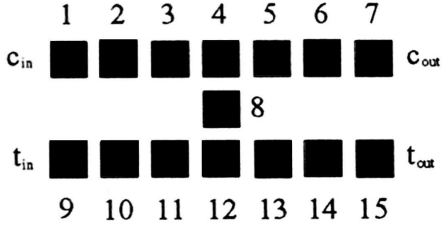}
    \end{center}
    \caption{cluster's structure for CNOT gate.}
\end{figure}

\section{Tabular notation}
To simplify the presentation of calculations we developed and
introduced a new notation, which we named ``tabular notation''.

Any cluster state is a sum of many tensorial products, in which
are written the state of all cluster qubits. We can show them as
items of a row in a table which has a number of columns equal to
qubits number plus one. The first column is named ``sign'', the
remaining ones give the qubit position in the cluster. Into the
column ``sign'' the algebraic sign of the tensorial products is
written and into the numbered columns the qubits' states. Any
tensorial product is thereby written into a line, and the cluster
state is represented by the whole table. With this notation, we
can calculate the cluster state more easily without mistakes.

\section{Entangled state formation}
Refering to fig.1, the cluster for the CNOT gate is divided in
subclusters formed by 1, 2, 3,  or 4 qubits. The entanglement
operator is applied to all subclusters. Then, the subclusters are
``repaired'' with the interaction of marginal qubits. In this way,
we can piece together the whole cluster. We use the following
subclusters:
\begin{enumerate} \item from qubit 1 to qubit 3; \item from qubit 4 to qubit
7; \item from quibt 9 to qubit 11; \item from qubit 12 to qubit
15; \item the qubit 8. \end{enumerate}

We show now the way the operator $S^{(C)}$ is applied. Initially,
we take two qubits, in the $\mid+>$ state, that is:

\begin{eqnarray}
\mid\Psi>_{2} & = & \mid+>_{1}\mid+>_{2}
\end{eqnarray}
and then apply to this state the operator $S^{(2)}$. So, the
entangled state:

\begin{eqnarray}
S^{(2)}\mid\Psi>_{2}&=& \mid0>_{1}\mid+>_{2} +
\mid1>_{1}\mid->_{2}.
\end{eqnarray}
is obtained.

After, we take three qubits in the $\mid+>$:

\begin{eqnarray}
\mid\Psi_{3}> & = & \mid+>_{1}\mid+>_{2}\mid+>_{3},
\end{eqnarray}
appling the operator $S^{(3)}$, one obtains:

\begin{eqnarray}
S^{(3)}\mid\Psi>_{3}& = & \mid+>_{1}\mid0>_{2}\mid+>_{3} \nonumber
\\ & + & \mid->_{1}\mid1>_{2}\mid->_{3}.
\end{eqnarray}

In the case of 4 qubits:

\begin{eqnarray}
\mid\Psi>_{4} & = & \mid+>_{1}\mid+>_{2}\mid+>_{3}\mid+>_{4}
\end{eqnarray}
the result is:

\begin{eqnarray}
S^{(4)}\mid\Psi>_{4} & = &
\mid0>_{1}\mid+>_{2}\mid0>_{3}\mid+>_{4} \nonumber \\ & + &
\mid0>_{1}\mid->_{2}\mid1>_{3}\mid->_{4} \nonumber \\ & + &
\mid1>_{1}\mid->_{2}\mid0>_{3}\mid+>_{4} \nonumber \\ & + &
\mid1>_{1}\mid+>_{2}\mid1>_{3}\mid->_{4}. \nonumber \\ & &
\end{eqnarray}
Whereas in the case of 5 qubits:

\begin{eqnarray}
\mid\Psi>_{1,2,3,4,5} =
\mid\psi_{in}>_{1}\mid+>_{2}\mid+>_{3}\mid+>_{4}\mid+>_{5}
\end{eqnarray}
the result is:

\begin{eqnarray}
S^{(5)}\mid\Psi>_{1,2,3,4,5} &=&
\mid\psi_{in}>_{1}\mid0>_{2}\mid+>_{3}\mid0>_{4}\mid+>_{5}
\nonumber \\& + &
\mid\psi_{in}>_{1}\mid0>_{2}\mid->_{3}\mid1>_{4}\mid->_{5}
\nonumber \\& + &
\mid\psi_{in}\ast>_{1}\mid1>_{2}\mid->_{3}\mid0>_{4}\mid+>_{5}
\nonumber \\& + &
\mid\psi_{in}\ast>_{1}\mid1>_{2}\mid+>_{3}\mid1>_{4}\mid->_{5}\nonumber
\\ & &
\end{eqnarray}
where the state $\mid\psi_{in}>$ and $\mid\psi_{in}\ast>$ are
linear combinations of the states $\mid 0>$ and $\mid 1>$:

\begin{eqnarray} \mid\psi_{in}> & = & a\mid0> +
b\mid1>
\end{eqnarray}

\begin{eqnarray} \mid\psi_{in}\ast> & = & a\mid0> -
b\mid1>.
\end{eqnarray}

The states obtained by us are different from the results published
by Raussendorf and Briegel \cite{2}, \cite{3}. Their result for
the cluster of 4 and 5 qubits  are respectivily:

\begin{eqnarray}
S^{(4)}\mid\Psi>_{4} & = &
\mid+>_{1}\mid0>_{2}\mid+>_{3}\mid0>_{4}\nonumber \\ &
+ & \mid+>_{1}\mid0>_{2}\mid->_{3}\mid1>_{4} \nonumber \\
& + & \mid->_{1}\mid1>_{2}\mid->_{3}\mid0>_{4} \nonumber \\ & + &
\mid->_{1}\mid1>_{2}\mid+>_{3}\mid1>_{4}.\nonumber \\ & &
\end{eqnarray}
\begin{eqnarray}
S^{(5)}\mid\Psi>_{1,2,3,4,5} & = &
\mid\psi_{in}>_{1}\mid0>_{2}\mid->_{3}\mid0>_{4}\mid->_{5} \nonumber \\
& - & \mid\psi_{in}>_{1}\mid0>_{2}\mid+>_{3}\mid1>_{4}\mid+>_{5} \nonumber \\
& - & \mid\psi_{in}\ast>_{1}\mid1>_{2}\mid+>_{3}\mid0>_{4}\mid->_{5} \nonumber \\
& + &
\mid\psi_{in}\ast>_{1}\mid1>_{2}\mid->_{3}\mid1>_{4}\mid+>_{5}.\nonumber
\\ & &
\end{eqnarray}
It is important to note that their results are not compatible
between them.

We used our results to calculate the cluster state for 15 qubits.
The relevant results are reported in appendix B. In this way, we
have obtained the expression of the cluster state for the one-way
CNOT gate.


\section{Measurements onto the cluster}
We name the input qubits $C_{I}(g)$, the output qubits $C_{O}(g)$,
and the measurements cluster $C_{M}(g)$. The implementation of
unitary transformation $C(CNOT)$ is strictly tied to the
fundamental theorem of the one-way quantum computation (see
appendix B).

We applied the fundamental theorem for the analysis of CNOT gate.
The qubits 1 and 8 give $C_{I}(g)$, the qubits 7 e 15 $C_{O}(g)$,
all others qubits $C_{M}(g)$. Let $\mid \phi>$ be a cluster state
on $C(CNOT)$ which satisfies the set of eigenvalue equations (10).
From these basic eigenvalue equations one gets:
\begin{widetext}
\begin{eqnarray}
\mid\phi> & = &
K^{(1)}K^{(3)}K^{(4)}K^{(5)}K^{(7)}K^{(8)}K^{(13)}K^{(15)}\mid\phi>
\nonumber \\
& = &
-{\sigma_{x}}^{(1)}{\sigma_{y}}^{(3)}{\sigma_{y}}^{(4)}{\sigma_{y}}^{(5)}{\sigma_{x}}^{(7)}{\sigma_{y}}^{(8)}{\sigma_{x}}^{(13)}{\sigma_{x}}^{(15)}\mid\phi>
\\
\mid\phi> & = & K^{(2)}K^{(3)}K^{(5)}K^{(6)}\mid\phi> \nonumber \\
& = &
{\sigma_{z}}^{(1)}{\sigma_{y}}^{(2)}{\sigma_{y}}^{(3)}{\sigma_{y}}^{(5)}{\sigma_{y}}^{(6)}{\sigma_{z}}^{(7)}\\
\mid\phi> & = & K^{(9)}K^{(11)}K^{(13)}K^{(15)}\mid\phi> \nonumber \\
& = &
{\sigma_{x}}^{(9)}{\sigma_{x}}^{(11)}{\sigma_{x}}^{(13)}{\sigma_{x}}^{(15)}\mid\phi>
\\
\mid\phi> & = &
K^{(5)}K^{(6)}K^{(8)}K^{(10)}K^{(12)}K^{(14)}\mid\phi> \nonumber
\\
& = &
{\sigma_{y}}^{(5)}{\sigma_{y}}^{(6)}{\sigma_{z}}^{(7)}{\sigma_{y}}^{(8)}{\sigma_{z}}^{(9)}{\sigma_{x}}^{(10)}{\sigma_{y}}^{(12)}{\sigma_{x}}^{(14)}{\sigma_{z}}^{15)}\mid\phi>.
\end{eqnarray}
\end{widetext}

If the qubits 10, 11, 13 and 14 are measured in the $\sigma_{x}$
eigenbasis and the qubits 2, 3, 4, 5, 6, 8 and 12 are measured in
the $\sigma_{y}$ eigenbasis, then the measurement results $s_{2},
s_{3}, s_{4}, s_{5}, s_{6}, s_{8}, s_{10}, s_{11}, s_{12}, s_{13},
s_{14} \in {\{0,1\}}$ are obtained. The cluster state eigenvalue
equations (28), (29), (30), (31) lead to the following eigenvalue
equations for the projected state:

\begin{eqnarray}
{\sigma_{x}}^{(1)}{\sigma_{x}}^{(7)}{\sigma_{x}}^{(15)}\mid\psi>
\nonumber\\
=  (-1)^{1+s_{3}+s_{4}+s_{5}+s_{8}+s_{13}}\mid\psi> \\
{\sigma_{z}}^{(1)}{\sigma_{z}}^{(7)}\mid\psi> \nonumber\\ =
(-1)^{s_{2}+s_{3}+s_{5}+s_{6}}\mid\psi> \\
{\sigma_{x}}^{(9)}{\sigma_{x}}^{(15)}\mid\psi>\nonumber \\ =
(-1)^{s_{11}+s_{13}}\mid\psi> \\
{\sigma_{z}}^{(7)}{\sigma_{z}}^{(9)}{\sigma_{z}}^{(15)}\mid\psi>
\nonumber
\\ =  (-1)^{s_{5}+s_{6}+s_{8}+s_{10}+s_{12}+s_{14}}\mid\psi>.
\end{eqnarray}

Therein, qubits 1 and 7 represent the input and output for the
control qubit and qubits 9 and 15 represent the input and output
for the target qubit. Writing the CNOT unitary operation on
control and target qubits $CNOT(c,t)$, we find:

\begin{eqnarray}
CNOT(c,t){\sigma_{x}}^{(c)}CNOT(c,t) & = &
{\sigma_{x}}^{(c)}{\sigma_{x}}^{(t)} \\
CNOT(c,t){\sigma_{z}}^{(c)}CNOT(c,t) & = & {\sigma_{z}}^{(c)} \\
CNOT(c,t){\sigma_{x}}^{(t)}CNOT(c,t) & = & {\sigma_{x}}^{(t)} \\
CNOT(c,t){\sigma_{z}}^{(t)}CNOT(c,t) & = &
{\sigma_{z}}^{(c)}{\sigma_{z}}^{(t)}.
\end{eqnarray}

Comparing these equations to the eigenvalue equations (32) to
(35), one clearly sees that measurements do indeed realize a CNOT
gate. Furthermore, after reading off the operator $U_{\sum}$ using
equations (A2) and (A4) and propagating the byproduct operators
through to the output side of the CNOT gate, one finds the
following expression for the byproduct operators of CNOT gate:

\begin{eqnarray}
U_{\sum,CNOT}=
{\sigma_{x}}^{(c)^{{\gamma_{x}}^{(c)}}}{\sigma_{x}}^{(t)^{{\gamma_{x}}^{(t)}}}{\sigma_{z}}^{(c)^{{\gamma_{z}}^{(c)}}}{\sigma_{z}}^{(t)^{{\gamma_{z}}^{(t)}}}
\end{eqnarray}
with

\begin{eqnarray}
{\gamma_{x}}^{(c)} & = & s_{2}+s_{3}+s_{5}+s_{6} \nonumber \\
{\gamma_{x}}^{(t)} & = & s_{2}+s_{3}+s_{8}+s_{10}+s_{12}+s_{14} \nonumber \\
{\gamma_{z}}^{(c)} & = &
s_{1}+s_{3}+s_{4}+s_{5}+s_{8}+s_{9}+s_{11}+
1 \nonumber \\
{\gamma_{z}}^{(t)} & = &  s_{9}+s_{11}+s_{13}.
\end{eqnarray}

The measurements set applied to the cluster is showed in fig.2.

\begin{figure}[h]
    \begin{center}
    \includegraphics[width=5cm]{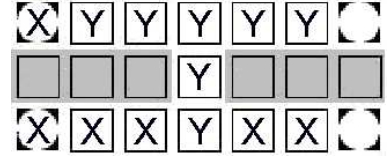}
    \end{center}
    \caption{measurements onto CNOT cluster.}
\end{figure}

\section{Conclusions} In this note we presented theoretical
results on the simulation of CNOT gate, based on the one-way
quantum computation model. The first complete simulation of CNOT
gate onto a cluster state is exhibited. Being the CNOT a universal
gate, its simulation makes valid the one-way model of quantum
computer.


\appendix
\section{Theorem}
The fundamental theorem of the one-way quantum computation is the
following \cite{3}. \vspace{2mm}\\Theor.:\\
\emph{be\\ $C(g) = C_{I}(g)\cup C_{M}(g)\cup C_{O}(g)$\\ with\\
$C_{I}(g)\cap C_{M}(g) = C_{I}(g)\cap C_{O}(g)= C_{M}(g)\cap
C_{O}(g)=\emptyset$\\ a cluster for the simulation of a gate
``g'', realizing the unitary trasformation U and
$\mid\phi>_{C(g)}$ the cluster state on the cluster $C(g)$. Ones
measures the qubits $C_{M}(g)$, that is projects the initial state
$\mid\phi>_{C(g)}$
in the new state \\
$\mid\psi>_{C(g)}={P_{\{s\}}}^{(C_{M}(g))}\mid\phi>_{C(g)}$. The
measuremnt pattern $\mathcal{M^{(C)}}$ which is applied to the
$C_{M}(g)$ qubits is a set of vectors:
\begin{eqnarray}
\mathcal{M^{(C)}}={\{\vec{r}_{a} \in S^{2}\mid a \in \mathcal{C}
\}}.
\end{eqnarray}
This vectors determine the basis for all measurements onto the
cluster.\\ Suppose, the new state $\mid\psi>_{C(g)}$ obeys the 2n
eigenvalue equations:
\begin{eqnarray}
{\sigma_{x}}^{(C_{I}(g),i)}(U{\sigma_{x}}^{(i)}U^{\dag})^{(C_{O}(g))}\mid\psi>_{C(g)}
 = \\ (-1)^{\lambda_{x,i}}\mid\psi>_{C(g)} \nonumber \\
{\sigma_{z}}^{(C_{I}(g),i)}(U{\sigma_{z}}^{(i)}U^{\dag})^{(C_{O}(g))}\mid\psi>_{C(g)}
 =\\  (-1)^{\lambda_{z,i}}\mid\psi>_{C(g)}
\end{eqnarray}
where $\lambda_{x,i}, \lambda_{z,i} \in {\{0,1\}}$ e $1\leq i\leq
n$ and ``n'' shows the number of logic qubits  which must be
processed. Then, the logic gate ``g'' which acts onto the initial
state $\mid\psi_{in}>$ is carried out onto the cluster $C(g)$
appling to the qubits $C_{M}(g)$ the projector's operator
${P_{\{s\}}}^{(C_{M}(g))}$ and to the qubits $C_{I}(g)$ the
operator $\sigma_{x}$.\\
In this way, the input and the output state in the simulation of
``g'' are related via:
\begin{eqnarray}
\mid\psi_{out}>=UU_{\sum}\mid\psi_{in}>
\end{eqnarray}
where $U_{\sum}$ is a byproduct operator given by:
\begin{eqnarray}
U_{\sum}=\bigotimes_{i =
1...n}({\sigma_{z}}^{[i]})^{s_{i}+\lambda_{x,i}}({\sigma_{x}}^{[i]})^{\lambda_{z,i}}.
\end{eqnarray}}

\section{Relevant results} The entangled state
$S^{(3)}\mid\Psi_{1,2,3}>$ is represented by:
\begin{table}[h!]
\begin{tabular}{|c|c|c|c|}
\hline sign&1&2&3 \\\hline \hline
+&$\psi_{in}$&0&+ \\
+&$\psi_{in}\ast$&1&-\\
\hline
\end{tabular}
\end{table}

The entangled state $S^{(3)}\mid\Psi_{9,10,,11}>$ is represented
by:
\begin{table}[h!]
\begin{tabular}{|c|c|c|c|}
\hline sign&9&10&11 \\\hline \hline
+&$\psi_{in}$&0&+ \\
+&$\psi_{in}\ast$&1&-\\
\hline
\end{tabular}
\end{table}

The entangled state $S^{(4)}\mid\Psi_{4,5,6,7}>$ is represented
by:
\begin{table}[h!]
\begin{tabular}{|c|c|c|c|c|}
\hline sign&4&5&6&7\\
\hline \hline $+$&$0$&$+$&$0$&$+$ \\
$+$&$0$&$-$&$1$&$-$\\
$+$&$1$&$-$&$0$&$+$\\
$+$&$1$&$+$&$1$&$-$\\
\hline
\end{tabular}
\end{table}

The entangled state $S^{(4)}\mid\Psi_{12,13,14,15}>$ is
represented by:
\begin{table}[h!]
\begin{tabular}{|c|c|c|c|c|}
\hline sign&$12$&$13$&$14$&$15$\\
\hline \hline $+$&$0$&$+$&$0$&$+$ \\
$+$&$0$&$-$&$1$&$-$\\
$+$&$1$&$-$&$0$&$+$\\
$+$&$1$&$+$&$1$&$-$\\
\hline
\end{tabular}
\end{table}

The entangled state $S^{(3\cdot4)}\mid\Psi_{1,2,3,4,5,6,7}>$ is
represented by:
\begin{table}[h!]
\begin{tabular}{|c|c|c|c|c|c|c|c|}
\hline sign&1&2&3&4&5&6&7 \\\hline \hline
+&$\psi_{in}$&0&+&0&+&0&+ \\+&$\psi_{in}$&0&+&0&-&1&-\\
+&$\psi_{in}$&0&-&1&-&0&+\\ +&$\psi_{in}$&0&-&1&+&1&-\\
+&$\psi_{in}\ast$&1&-&0&+&0&+\\
+&$\psi_{in}\ast$&1&-&0&-&1&-\\
+&$\psi_{in}\ast$&1&+&1&-&0&+\\
+&$\psi_{in}\ast$&1&+&1&+&1&-\\
\hline
\end{tabular}
\end{table}
\newpage
The entangled state
$S^{(11\cdot12)}\mid\Psi_{9,10,11,12,13,14,15}>$ is represented
by:
\begin{table}[h!]
\begin{tabular}{|c|c|c|c|c|c|c|c|}
\hline sign&9&10&11&12&13&14&15\\
\hline \hline
+&$\psi_{in}$&0&+&0&+&0&+\\
+&$\psi_{in}$&0&+&0&-&1&-\\
+&$\psi_{in}$&0&-&1&-&0&+\\
+&$\psi_{in}$&0&-&1&+&1&-\\
+&$\psi_{in}\ast$&1&-&0&+&0&+\\
+&$\psi_{in}\ast$&1&-&0&-&1&-\\
+&$\psi_{in}\ast$&1&+&1&-&0&+\\
+&$\psi_{in}\ast$&1&+&1&+&1&-\\
\hline
\end{tabular}
\end{table}
\vspace{1.5cm} \\

For the state $S^{(4\cdot8)}\mid\Psi_{1,2,3,4,5,6,7,8}>$ the
result is:
\begin{table}[h!]
\begin{tabular}{|c|c|c|c|c|c|c|c|c|}
\hline
sign&1&2&3&4&5&6&7&8 \\
\hline \hline
+&$\psi_{in}$&0&+&0&+&0&+&0\\
+&$\psi_{in}$&0&+&0&+&0&+&1\\
+&$\psi_{in}$&0&+&0&-&1&-&0\\
+&$\psi_{in}$&0&+&0&-&1&-&1\\
+&$\psi_{in}$&0&-&1&-&0&+&0\\
-&$\psi_{in}$&0&-&1&-&0&+&1\\
+&$\psi_{in}$&0&-&1&+&1&-&0\\
-&$\psi_{in}$&0&-&1&+&1&-&1\\
+&$\psi_{in}\ast$&1&-&0&+&0&+&0\\
+&$\psi_{in}\ast$&1&-&0&+&0&+&1\\
+&$\psi_{in}\ast$&1&-&0&-&1&-&0\\
+&$\psi_{in}\ast$&1&-&0&-&1&-&1\\
+&$\psi_{in}\ast$&1&+&1&-&0&+&0\\
-&$\psi_{in}\ast$&1&+&1&-&0&+&1\\
+&$\psi_{in}\ast$&1&+&1&+&1&-&0\\
-&$\psi_{in}\ast$&1&+&1&+&1&-&1\\
\hline
\end{tabular}
\end{table}

The last result is the state
$S^{(8\cdot12)}\mid\Psi_{1,...,15}>$.The result is broken in two
parts for typographical reasons:
\newpage

\begin{table}[h!]
\begin{tabular}{|c|c|c|c|c|c|c|c|c|c|c|c|c|c|c|c|}
\hline
sign&1&2&3&4&5&6&7&8&9&10&11&12&13&14&15\\
\hline \hline
+&$\psi_{in}$&0&+&0&+&0&+&0&$\psi_{in}$&0&+&0&+&0&+\\
+&$\psi_{in}$&0&+&0&+&0&+&0&$\psi_{in}$&0&+&0&-&1&-\\
+&$\psi_{in}$&0&+&0&+&0&+&0&$\psi_{in}$&0&-&1&-&0&+\\
+&$\psi_{in}$&0&+&0&+&0&+&0&$\psi_{in}$&0&-&1&+&1&-\\
+&$\psi_{in}$&0&+&0&+&0&+&0&$\psi_{in}\ast$&1&-&0&+&0&+\\
+&$\psi_{in}$&0&+&0&+&0&+&0&$\psi_{in}\ast$&1&-&0&-&1&-\\
+&$\psi_{in}$&0&+&0&+&0&+&0&$\psi_{in}\ast$&1&+&1&-&0&+\\
+&$\psi_{in}$&0&+&0&+&0&+&0&$\psi_{in}\ast$&1&+&1&+&1&-\\
+&$\psi_{in}$&0&+&0&+&0&+&1&$\psi_{in}$&0&+&0&+&0&+\\
+&$\psi_{in}$&0&+&0&+&0&+&1&$\psi_{in}$&0&+&0&-&1&-\\
-&$\psi_{in}$&0&+&0&+&0&+&1&$\psi_{in}$&0&-&1&-&0&+\\
-&$\psi_{in}$&0&+&0&+&0&+&1&$\psi_{in}$&0&-&1&+&1&-\\
+&$\psi_{in}$&0&+&0&+&0&+&1&$\psi_{in}\ast$&1&-&0&+&0&+\\
+&$\psi_{in}$&0&+&0&+&0&+&1&$\psi_{in}\ast$&1&-&0&-&1&-\\
-&$\psi_{in}$&0&+&0&+&0&+&1&$\psi_{in}\ast$&1&+&1&-&0&+\\
-&$\psi_{in}$&0&+&0&+&0&+&1&$\psi_{in}\ast$&1&+&1&+&1&-\\
+&$\psi_{in}$&0&+&0&-&1&-&0&$\psi_{in}$&0&+&0&+&0&+\\
+&$\psi_{in}$&0&+&0&-&1&-&0&$\psi_{in}$&0&+&0&-&1&-\\
+&$\psi_{in}$&0&+&0&-&1&-&0&$\psi_{in}$&0&-&1&-&0&+\\
+&$\psi_{in}$&0&+&0&-&1&-&0&$\psi_{in}$&0&-&1&+&1&-\\
+&$\psi_{in}$&0&+&0&-&1&-&0&$\psi_{in}\ast$&1&-&0&+&0&+\\
+&$\psi_{in}$&0&+&0&-&1&-&0&$\psi_{in}\ast$&1&-&0&-&1&-\\
+&$\psi_{in}$&0&+&0&-&1&-&0&$\psi_{in}\ast$&1&+&1&-&0&+\\
+&$\psi_{in}$&0&+&0&-&1&-&0&$\psi_{in}\ast$&1&+&1&+&1&-\\
+&$\psi_{in}$&0&+&0&-&1&-&1&$\psi_{in}$&0&+&0&+&0&+\\
+&$\psi_{in}$&0&+&0&-&1&-&1&$\psi_{in}$&0&+&0&-&1&-\\
-&$\psi_{in}$&0&+&0&-&1&-&1&$\psi_{in}$&0&-&1&-&0&+\\
-&$\psi_{in}$&0&+&0&-&1&-&1&$\psi_{in}$&0&-&1&+&1&-\\
+&$\psi_{in}$&0&+&0&-&1&-&1&$\psi_{in}\ast$&1&-&0&+&0&+\\
+&$\psi_{in}$&0&+&0&-&1&-&1&$\psi_{in}\ast$&1&-&0&-&1&-\\
-&$\psi_{in}$&0&+&0&-&1&-&1&$\psi_{in}\ast$&1&+&1&-&0&+\\
-&$\psi_{in}$&0&+&0&-&1&-&1&$\psi_{in}\ast$&1&+&1&+&1&-\\
+&$\psi_{in}$&0&-&1&-&0&+&0&$\psi_{in}$&0&+&0&+&0&+\\
+&$\psi_{in}$&0&-&1&-&0&+&0&$\psi_{in}$&0&+&0&-&1&-\\
+&$\psi_{in}$&0&-&1&-&0&+&0&$\psi_{in}$&0&-&1&-&0&+\\
+&$\psi_{in}$&0&-&1&-&0&+&0&$\psi_{in}$&0&-&1&+&1&-\\
+&$\psi_{in}$&0&-&1&-&0&+&0&$\psi_{in}\ast$&1&-&0&+&0&+\\
+&$\psi_{in}$&0&-&1&-&0&+&0&$\psi_{in}\ast$&1&-&0&-&1&-\\
+&$\psi_{in}$&0&-&1&-&0&+&0&$\psi_{in}\ast$&1&+&1&-&0&+\\
+&$\psi_{in}$&0&-&1&-&0&+&0&$\psi_{in}\ast$&1&+&1&+&1&-\\
-&$\psi_{in}$&0&-&1&-&0&+&1&$\psi_{in}$&0&+&0&+&0&+\\
-&$\psi_{in}$&0&-&1&-&0&+&1&$\psi_{in}$&0&+&0&-&1&-\\
+&$\psi_{in}$&0&-&1&-&0&+&1&$\psi_{in}$&0&-&1&-&0&+\\
+&$\psi_{in}$&0&-&1&-&0&+&1&$\psi_{in}$&0&-&1&+&1&-\\
-&$\psi_{in}$&0&-&1&-&0&+&1&$\psi_{in}\ast$&1&-&0&+&0&+\\
-&$\psi_{in}$&0&-&1&-&0&+&1&$\psi_{in}\ast$&1&-&0&-&1&-\\
+&$\psi_{in}$&0&-&1&-&0&+&1&$\psi_{in}\ast$&1&+&1&-&0&+\\
+&$\psi_{in}$&0&-&1&-&0&+&1&$\psi_{in}\ast$&1&+&1&+&1&-\\
+&$\psi_{in}$&0&-&1&+&1&-&0&$\psi_{in}$&0&+&0&+&0&+\\
+&$\psi_{in}$&0&-&1&+&1&-&0&$\psi_{in}$&0&+&0&-&1&-\\
+&$\psi_{in}$&0&-&1&+&1&-&0&$\psi_{in}$&0&-&1&-&0&+\\
+&$\psi_{in}$&0&-&1&+&1&-&0&$\psi_{in}$&0&-&1&+&1&-\\
+&$\psi_{in}$&0&-&1&+&1&-&0&$\psi_{in}\ast$&1&-&0&+&0&+\\
+&$\psi_{in}$&0&-&1&+&1&-&0&$\psi_{in}\ast$&1&-&0&-&1&-\\
+&$\psi_{in}$&0&-&1&+&1&-&0&$\psi_{in}\ast$&1&+&1&-&0&+\\
+&$\psi_{in}$&0&-&1&+&1&-&0&$\psi_{in}\ast$&1&+&1&+&1&-\\
-&$\psi_{in}$&0&-&1&+&1&-&1&$\psi_{in}$&0&+&0&+&0&+\\
-&$\psi_{in}$&0&-&1&+&1&-&1&$\psi_{in}$&0&+&0&-&1&-\\
+&$\psi_{in}$&0&-&1&+&1&-&1&$\psi_{in}$&0&-&1&-&0&+\\
+&$\psi_{in}$&0&-&1&+&1&-&1&$\psi_{in}$&0&-&1&+&1&-\\
-&$\psi_{in}$&0&-&1&+&1&-&1&$\psi_{in}\ast$&1&-&0&+&0&+\\
-&$\psi_{in}$&0&-&1&+&1&-&1&$\psi_{in}\ast$&1&-&0&-&1&-\\
+&$\psi_{in}$&0&-&1&+&1&-&1&$\psi_{in}\ast$&1&+&1&-&0&+\\
+&$\psi_{in}$&0&-&1&+&1&-&1&$\psi_{in}\ast$&1&+&1&+&1&-\\
\hline
\end{tabular}
\end{table}

\begin{table}[h!]
\begin{tabular}{|c|c|c|c|c|c|c|c|c|c|c|c|c|c|c|c|}
\hline
+&$\psi_{in}\ast$&1&-&0&+&0&+&0&$\psi_{in}$&0&+&0&+&0&+\\
+&$\psi_{in}\ast$&1&-&0&+&0&+&0&$\psi_{in}$&0&+&0&-&1&-\\
+&$\psi_{in}\ast$&1&-&0&+&0&+&0&$\psi_{in}$&0&-&1&-&0&+\\
+&$\psi_{in}\ast$&1&-&0&+&0&+&0&$\psi_{in}$&0&-&1&+&1&-\\
+&$\psi_{in}\ast$&1&-&0&+&0&+&0&$\psi_{in}\ast$&1&-&0&+&0&+\\
+&$\psi_{in}\ast$&1&-&0&+&0&+&0&$\psi_{in}\ast$&1&-&0&-&1&-\\
+&$\psi_{in}\ast$&1&-&0&+&0&+&0&$\psi_{in}\ast$&1&+&1&-&0&+\\
+&$\psi_{in}\ast$&1&-&0&+&0&+&0&$\psi_{in}\ast$&1&+&1&+&1&-\\
+&$\psi_{in}\ast$&1&-&0&+&0&+&1&$\psi_{in}$&0&+&0&+&0&+\\
+&$\psi_{in}\ast$&1&-&0&+&0&+&1&$\psi_{in}$&0&+&0&-&1&-\\
+&$\psi_{in}\ast$&1&-&0&+&0&+&1&$\psi_{in}$&0&-&1&-&0&+\\
+&$\psi_{in}\ast$&1&-&0&+&0&+&1&$\psi_{in}$&0&-&1&+&1&-\\
+&$\psi_{in}\ast$&1&-&0&+&0&+&1&$\psi_{in}\ast$&1&-&0&+&0&+\\
+&$\psi_{in}\ast$&1&-&0&+&0&+&1&$\psi_{in}\ast$&1&-&0&-&1&-\\
+&$\psi_{in}\ast$&1&-&0&+&0&+&1&$\psi_{in}\ast$&1&+&1&-&0&+\\
+&$\psi_{in}\ast$&1&-&0&+&0&+&1&$\psi_{in}\ast$&1&+&1&+&1&-\\
+&$\psi_{in}\ast$&1&-&0&-&1&-&0&$\psi_{in}$&0&+&0&+&0&+\\
+&$\psi_{in}\ast$&1&-&0&-&1&-&0&$\psi_{in}$&0&+&0&-&1&-\\
+&$\psi_{in}\ast$&1&-&0&-&1&-&0&$\psi_{in}$&0&-&1&-&0&+\\
+&$\psi_{in}\ast$&1&-&0&-&1&-&0&$\psi_{in}$&0&-&1&+&1&-\\
+&$\psi_{in}\ast$&1&-&0&-&1&-&0&$\psi_{in}\ast$&1&-&0&+&0&+\\
+&$\psi_{in}\ast$&1&-&0&-&1&-&0&$\psi_{in}\ast$&1&-&0&-&1&-\\
+&$\psi_{in}\ast$&1&-&0&-&1&-&0&$\psi_{in}\ast$&1&+&1&-&0&+\\
+&$\psi_{in}\ast$&1&-&0&-&1&-&0&$\psi_{in}\ast$&1&+&1&+&1&-\\
+&$\psi_{in}\ast$&1&-&0&-&1&-&1&$\psi_{in}$&0&+&0&+&0&+\\
+&$\psi_{in}\ast$&1&-&0&-&1&-&1&$\psi_{in}$&0&+&0&-&1&-\\
-&$\psi_{in}\ast$&1&-&0&-&1&-&1&$\psi_{in}$&0&-&1&-&0&+\\
-&$\psi_{in}\ast$&1&-&0&-&1&-&1&$\psi_{in}$&0&-&1&+&1&-\\
+&$\psi_{in}\ast$&1&-&0&-&1&-&1&$\psi_{in}\ast$&1&-&0&+&0&+\\
+&$\psi_{in}\ast$&1&-&0&-&1&-&1&$\psi_{in}\ast$&1&-&0&-&1&-\\
-&$\psi_{in}\ast$&1&-&0&-&1&-&1&$\psi_{in}\ast$&1&+&1&-&0&+\\
-&$\psi_{in}\ast$&1&-&0&-&1&-&1&$\psi_{in}\ast$&1&+&1&+&1&-\\
+&$\psi_{in}\ast$&1&+&1&-&0&+&0&$\psi_{in}$&0&+&0&+&0&+\\
+&$\psi_{in}\ast$&1&+&1&-&0&+&0&$\psi_{in}$&0&+&0&-&1&-\\
+&$\psi_{in}\ast$&1&+&1&-&0&+&0&$\psi_{in}$&0&-&1&-&0&+\\
+&$\psi_{in}\ast$&1&+&1&-&0&+&0&$\psi_{in}$&0&-&1&+&1&-\\
+&$\psi_{in}\ast$&1&+&1&-&0&+&0&$\psi_{in}\ast$&1&-&0&+&0&+\\
+&$\psi_{in}\ast$&1&+&1&-&0&+&0&$\psi_{in}\ast$&1&-&0&-&1&-\\
+&$\psi_{in}\ast$&1&+&1&-&0&+&0&$\psi_{in}\ast$&1&+&1&-&0&+\\
+&$\psi_{in}\ast$&1&+&1&-&0&+&0&$\psi_{in}\ast$&1&+&1&+&1&-\\
-&$\psi_{in}\ast$&1&+&1&-&0&+&1&$\psi_{in}$&0&+&0&+&0&+\\
-&$\psi_{in}\ast$&1&+&1&-&0&+&1&$\psi_{in}$&0&+&0&-&1&-\\
+&$\psi_{in}\ast$&1&+&1&-&0&+&1&$\psi_{in}$&0&-&1&-&0&+\\
+&$\psi_{in}\ast$&1&+&1&-&0&+&1&$\psi_{in}$&0&-&1&+&1&-\\
-&$\psi_{in}\ast$&1&+&1&-&0&+&1&$\psi_{in}\ast$&1&-&0&+&0&+\\
-&$\psi_{in}\ast$&1&+&1&-&0&+&1&$\psi_{in}\ast$&1&-&0&-&1&-\\
+&$\psi_{in}\ast$&1&+&1&-&0&+&1&$\psi_{in}\ast$&1&+&1&-&0&+\\
+&$\psi_{in}\ast$&1&+&1&-&0&+&1&$\psi_{in}\ast$&1&+&1&+&1&-\\
+&$\psi_{in}\ast$&1&+&1&+&1&-&0&$\psi_{in}$&0&+&0&+&0&+\\
+&$\psi_{in}\ast$&1&+&1&+&1&-&0&$\psi_{in}$&0&+&0&-&1&-\\
+&$\psi_{in}\ast$&1&+&1&+&1&-&0&$\psi_{in}$&0&-&1&-&0&+\\
+&$\psi_{in}\ast$&1&+&1&+&1&-&0&$\psi_{in}$&0&-&1&+&1&-\\
+&$\psi_{in}\ast$&1&+&1&+&1&-&0&$\psi_{in}\ast$&1&-&0&+&0&+\\
+&$\psi_{in}\ast$&1&+&1&+&1&-&0&$\psi_{in}\ast$&1&-&0&-&1&-\\
+&$\psi_{in}\ast$&1&+&1&+&1&-&0&$\psi_{in}\ast$&1&+&1&-&0&+\\
+&$\psi_{in}\ast$&1&+&1&+&1&-&0&$\psi_{in}\ast$&1&+&1&+&1&-\\
-&$\psi_{in}\ast$&1&+&1&+&1&-&1&$\psi_{in}$&0&+&0&+&0&+\\
-&$\psi_{in}\ast$&1&+&1&+&1&-&1&$\psi_{in}$&0&+&0&-&1&-\\
+&$\psi_{in}\ast$&1&+&1&+&1&-&1&$\psi_{in}$&0&-&1&-&0&+\\
+&$\psi_{in}\ast$&1&+&1&+&1&-&1&$\psi_{in}$&0&-&1&+&1&-\\
-&$\psi_{in}\ast$&1&+&1&+&1&-&1&$\psi_{in}\ast$&1&-&0&+&0&+\\
-&$\psi_{in}\ast$&1&+&1&+&1&-&1&$\psi_{in}\ast$&1&-&0&-&1&-\\
+&$\psi_{in}\ast$&1&+&1&+&1&-&1&$\psi_{in}\ast$&1&+&1&-&0&+\\
+&$\psi_{in}\ast$&1&+&1&+&1&-&1&$\psi_{in}\ast$&1&+&1&+&1&-\\
\hline
\end{tabular}
\end{table}

\par\noindent



\begin{thebibliography}{5}
\bibitem{1} M. A. Nielsen, I. L. Chuang, \textit{Quantum Computation and Quantum Information}, Cambridge University
Press.
\bibitem{2} R. Raussendorf, H. J. Briegel, \textit{A One-Way Quantum Computer}, Physical Review Letters, vol. 86, No. 22, pag. 5188-5191, 2001.
\bibitem{3} R. Raussendorf, D. E. Browne, H. J. Briegel, \textit{Measurement-based quantum computation on cluster stastes}, arXiv: quantum-ph/0301052 v2, 2005.
\end{thebibliography}
\end{document}